\documentclass[aps,preprint]{revtex4}
\usepackage{amsfonts}
\usepackage{amsmath}
\usepackage{amssymb}
\usepackage{subfigure}
\usepackage{graphicx}
\usepackage{epstopdf}
\usepackage{float}
\begin{document}

\title{Test the Weak Cosmic Censorship Conjecture in Torus-Like Black Hole under Charged Scalar Field}
\author{Wei Hong$^{a}$}
\email{thphysics_weihong@stu.scu.edu.cn}
\author{Benrong Mu$^{b}$}
\email{benrongmu@cdutcm.edu.cn}
\author{Jun Tao$^{a}$}
\email{taojun@scu.edu.cn}
\affiliation{$^{a}$Center for Theoretical Physics, College of Physics, Sichuan University, Chengdu, 610065, China}
\affiliation{$^{b}$Physics Teaching and Research Section, College of Medical Technology, Chengdu University of Traditional Chinese Medicine, Chengdu 611137, China}

\begin{abstract}
We investigate weak cosmic censorship conjecture in charged torus-like black hole by the complex scalar field scattering. Using the relation between the conserved quantities of a black hole and the scalar field, we can calculate the change of the energy and charge within the infinitesimal time. The change of the enthalpy is connected to the change of energy, then we use those results to test  whether the first law, the second law as well as the weak cosmic censorship conjecture are valid. In the normal phase space, the first law of thermodynamics and the weak cosmic censorship conjecture are valid, and the second law of thermodynamics is not violated. For the specific black hole under scalar field scattering we consider, in the extend phase space, the first law of thermodynamics and the weak cosmic censorship conjecture are valid. However, the second law of thermodynamics is violated when the black hole's initial charge reaches a certain value.
\end{abstract}
\maketitle
%\tableofcontents

\section{Introduction}
As one of the most successful predictions of general relativity, black holes have received much attention since the first spherically symmetric vacuum solution to Einstein's field equation became known. With the first picture of black hole \cite{Akiyama:2019cqa,Akiyama:2019brx,Akiyama:2019sww,Akiyama:2019bqs,Akiyama:2019fyp,Akiyama:2019eap}, the study of black holes reached a climax again. To this day, people have carried out a lot of researches on black holes. Compared with the universal thermodynamic system, people have gradually established four laws of black hole thermodynamics \cite{Hawking:1976de,Bekenstein:1974ax,Bardeen:1973gs}. Beckenstein investigated whether the second law of thermodynamics could hold for black hole systems. He assumed that the second law of thermodynamics should be universal, from the point of view of information theory, that a black hole should have an entropy proportional to the area of its event horizon, and Hawking determined this coefficient and put Beckenstein's black hole entropy on the basis of thermodynamics, which is called Beckenstein-Hawking entropy \cite{Bekenstein:1973ur}. When considering quantum field theory into a black hole, Hawking found that a black hole radiates matter in the form of thermal radiation, the temperature of which is proportional to its surface gravity \cite{Hawking:1974sw}.

Over a long period of time, the study of the thermodynamic properties of black holes has become a fairly mature subject, especially in the field of phase transition and phase structure, people have done a lot of work. Hawking and Page's research shows that there is a first-order phase transition between the Schwarzschild AdS black hole and the thermal AdS space \cite{Hawking:1982dh}. In the context of AdS/CFT correspondence \cite{Witten:1998zw,Halperin:1998vy,Rey:1998ik,Henningson:1998cd}, this phase transition is later understood as confinement/deconfinement phase  transition. The AdS/CFT correspondence successfully solved the black hole information to some extent because it can show how the time evolution of the black hole can follow quantum mechanics to some extent. It is indeed feasible to consider black holes with the content of the AdS/CFT correspondence. Any such black hole corresponds to a series of particles located at the edge of the anti-de Sitter space. These particles normally follow the rules of quantum mechanics, especially the positive time evolution, so the black hole must also conform to the positive time evolution and obey the rules of quantum mechanics. In recent years, phase transition, phase structure, $P-V$ criticality and thermodynamics of various black holes in the context of extended phase space thermodynamics have been studied. In those studies, cosmological constants are interpreted as thermodynamic pressure \cite{Kastor:2009wy,Dolan:2011xt,Gunasekaran:2012dq,Wei:2012ui,Cai:2013qga,Xu:2014kwa,Frassino:2014pha,Dehghani:2014caa,Hennigar:2015esa,Wang:2018xdz}.

Penrose's weak cosmic censorship conjecture (WCCC) asserts that singularities are always hidden in the horizon of any actual physical process, rather than in the traditional initial conditions, the observer does not see it in the future null infinity \cite{Penrose:1969pc}. Since the physical behaviour of the singularity is unknown, if the singularity can be observed by other parts of time and space, the causal relationship will break and physics may lose the ability to predict. According to Penrose-Hawking singularity theorems, singularity is inevitable in the case of physical meaning, which makes WCCC inevitable. In addition, if the naked singularity does not exist, then the universe will become deterministic, which means that it is possible to speculate on the entire evolution of the universe based only on the state of the universe at a certain moment. To verify the correctness of the WCCC, Wald tried to overcharge or overspin the extreme Kerr-Newman black hole by placing the test particles in a black hole \cite{Wild.R}. However, for particles with sufficient charge or angular momentum to overcharge or overspin the black hole, the black hole cannot capture it due to electromagnetic or centrifugal repulsive forces. On the other hand, it has been found that the absorption of particles can overcharge/overspin the near-extreme charged/rotating black hole \cite{Hubeny:1998ga,Jacobson:2009kt,Saa:2011wq,Rocha:2014jma,BouhmadiLopez:2010vc,Gao:2012ca,Duztas:2016xfg}. Recently, Sorce and Wald have suggested a new version of the gedanken experiments to overspin or overcharge the Kerr-Newman black holes in Einstein-Maxwell gravity. Following their setup, there are already some works here \cite{He:2019mqy,Wang:2019bml,Jiang:2019vww}. Relative to particle absorption, our job is to place black holes near the scalar field and scatter the scalar field to see if this happens. Their validity has been tested in various black holes due to the lack of general evidence of WCCC.  Here are some articles on validity test: black hole absorbing particles \cite{Chen:2019pdj,Isoyama:2011ea,Mu:2019bim,Wang:2019dzl,Zeng:2019hux,Gwak:2017icn}, scattering of black holes under scalar fields \cite{Gwak:2019asi,Chen:2019nsr,Chen:2018yah,Gwak:2018akg,Hong:2019yiz,Bai:2020ieh,Gwak:2019rcz}, and other types \cite{Yu:2018eqq,Crisford:2017gsb,Husain:2017cmj,Duztas:2013wua,Hod:2013vj,Richartz:2008xm,Eperon:2019viw,Rocha:2011wp,Gim:2018axz}.

In this paper, we study the thermodynamics and the WCCC of a special black hole with complex field scattering in the normal and extended phase space, considering thermodynamic pressure and volume or not. This special black hole is the torus-like black hole \cite{Huang:1995zb}, which has a different space-time topology of $S\times S\times M^2$ than other black holes. Other work\cite{Han:2019kjr} has been done on the metric of the same black hole, except that they do it when the black hole absorbs scalar particles, and we do it under the scalar field scatters. Moreover, we inserted four images at specific locations in the article to illustrate the relevant contents. The structure of this paper is as follows. In the section II, we calculate the change of energy and electric charge of the black hole over a period of time. In the section III, we study the thermodynamics and WCCC of this black holes in normal phase space. In the section IV, we study the thermodynamics and WCCC of this black holes in extended phase space. In section V, we summarize our results.

\section{Variation of the torus-like black hole's energy and charge}
The charged torus-like black hole solution has a form \cite{Huang:1995zb,Han:2019kjr}
\begin{equation}
d s^{2}=-f(r)dt^{2}+\frac{dr^{2}}{f(r)}+r^{2}(d \theta ^{2}+d\psi^{2}),
\end{equation}
where
\begin{equation}
f(r)=-\frac{\Lambda r^{2}}{3}-\frac{2 M}{\pi r}+\frac{4 Q^{2}}{\pi r^{2}},
\end{equation}
here $M$ and $Q$ are mass and electric charge of the black hole. For $\Lambda <0$, the black hole can find it's horizon $r_{\pm}$ via solving the metric, which satisfy \cite{Sharif:2012se}
\begin{equation}
-\frac{\Lambda r_{\pm}^{2}}{3}-\frac{2 M}{\pi r_{\pm}}+\frac{4 Q^{2}}{\pi r_{\pm}^{2}}=0,\label{e3}
\end{equation}
where $r_{+}$ and $r_{-}$ are the outer and inner horizons of the black hole. The electromagnetic potential of the black hole is
\begin{equation}
\varphi= \frac{4 Q}{r_+}.\label{e4}
\end{equation}
The action of a complex scalar field in the fixed torus-like gravitational and electromagnetic fields is
\begin{equation}
\mathcal{S}=-\frac{1}{2} \int \sqrt{-g}\left[\left(\partial^{\mu}-i q A^{\mu}\right) \Psi^{*}\left(\partial_{\mu}+i q A_{\mu}\right) \Psi-m^{2} \Psi^{*} \Psi\right] d^{4} x,\label{e5}
\end{equation}
where $A_{\mu}$ is the electromagnetic potential, $m$ is the mass, $q$ is the charge, $\Psi$ represents the
wave function, and its conjugate is $\Psi^{*}$. Now we  investigate the dynamical of the charged complex scalar field. The field equation obtained from the action satisfies
\begin{equation}
\left(\nabla^{\mu}-i q A^{\mu}\right)\left(\nabla_{\mu}-i q A_{\mu}\right) \Psi-m^{2} \Psi=0.\label{e6}
\end{equation}
To solve this wave function, we carry out a separation of variables
\begin{equation}
\Psi=e^{-i \omega t} R(r) \Phi(\theta, \phi),\label{e7}
\end{equation}
where $\omega$ is the energy of the particle, and $\Phi(\theta, \phi)$ is the scalar spherical harmonics. We put Eq. (\ref{e7}) into Eq. (\ref{e6}) and obtain the radial wave function
\begin{equation}
R(r)=e^{ \pm i(\omega-4q Q / r) r_{*}},\label{e8}
\end{equation}
where $d r_{*}=\frac{1}{f} d r, r_{*}$ is a function of $r$ and $+/-$ represent the solution of the outgoing/ingoing radial wave. Thus, we can test thermodynamics and discuss the validity of the weak cosmic censorship conjecture by the scattering of ingoing waves at the event horizon. For this reason, we mainly concern the ingoing wave function in this paper. 

From Eq. (\ref{e5}), the energy-momentum tensor can be obtained as follow \cite{Chen:2019nsr,Gwak:2019asi,Semiz:2005gs,Toth:2011ab}
\begin{equation}
T_{\nu}^{\mu}=\frac{1}{2}\left[\left(\partial^{\mu}-i q A^{\mu}\right) \Psi^{*} \partial_{\nu} \Psi+\left(\partial^{\mu}+i q A^{\mu}\right) \Psi \partial_{\nu} \Psi^{*}\right]+\delta_{\nu}^{\mu} \mathcal{L}.
\end{equation}
The energy flux is produced by combining the ingoing wave function and its conjugate with the energy-momentum tensor \cite{Gwak:2019asi}
\begin{equation}
\frac{d E}{d t}=\int T_{t}^{r} \sqrt{-g} d \theta d \phi=\omega(\omega-q \varphi) r_{+}^{2}.\label{e10}
\end{equation}
The electric current is obtained from the Eq.  (\ref{e5})
\begin{equation}
j^{\mu}=\frac{\partial \mathcal{L}}{\partial A_{\mu}}=-\frac{1}{2} i q\left[\Psi^{*}\left(\partial^{\mu}+i q A^{\mu}\right) \Psi-\Psi\left(\partial^{\mu}-i q A^{\mu}\right) \Psi^{*}\right].
\end{equation}
For the ingoing wave in Eq.  (\ref{e8}), the charge flux \cite{Gwak:2019asi} is
\begin{equation}
\frac{d Q}{d t}=-\int j^{r} \sqrt{-g} d \theta d \varphi=q(\omega-q \varphi) r_{+}^{2}.\label{e12}
\end{equation}
When scattering complex scalar field, due to the conservation of energy and charge, the decrease of energy and charge in scalar fields is equal to the increase of energy and charge in black hole. From Eqs. (\ref{e10}) and (\ref{e12}), the transferred energy and charge within the certain time interval are
\begin{equation}
d U=d E=\omega(\omega-q \varphi) r_{+}^{2} d t, \quad d Q=q(\omega-q \varphi) r_{+}^{2} d t\label{e13}
\end{equation}
respectively. Since the transfer of energy and charge is small, time $dt$ must also be very small. The increase or decrease of $dU$ and $dQ$ depends on the relationship between $\omega$ and $q\varphi $.

\section{The thermodynamics and WCCC of the torus-like black hole in normal phase space}
Now we turn to the thermodynamics of torus-like black hole in normal phase space. According to the definition of surface gravity, the Hawking temperature of the black hole can be written as
\begin{equation}
T=\frac{f'(r_+)}{4\pi}.\label{e14}
\end{equation}
By using the relation between the Beckenstein-Hawking entropy and the surface area of event horizon, the entropy of a black hole can be written as \cite{Han:2019kjr}
\begin{equation}
S=\pi^{2} r_{+}^{2}.\label{e15}
\end{equation}

In the normal phase space, the initial state of the black hole is represented by $\left(M, Q, r_{+}\right)$, while the final state is represented by $\left(M+d M, Q+d Q, r_{+}+d r_{+}\right)$. The radius' variation can be obtained from variation of the metric component $f\left(r_{+}\right)$. For the initial state $\left(M, Q, r_{+}\right)$, satisfies
\begin{equation}
f\left(M, Q, r_{+}\right)=0.\label{e16}
\end{equation}
We assume that the final state of the black hole is still satisfied
\begin{equation}
f\left(M+d M, Q+d Q, r_{+}+d r_{+}\right)=0.\label{e17}
\end{equation}
The functions $f\left(M+d M, Q+d Q, r_{+}+d r_{+}\right)$ and $f\left(M, Q, r_{+}\right)$ satisfy the following relation
\begin{equation}
\begin{aligned} & f\left(M+d M, Q+d Q, r_{+}+d r_{+}\right) \\=& f\left(M, Q, r_{+}\right)+\left.\frac{\partial f}{\partial M}\right|_{r=r_{+}} d M+\left.\frac{\partial f}{\partial Q}\right|_{r=r_{+}} d Q+\left.\frac{\partial f}{\partial r_{+}}\right|_{r=r_{+}} d r_{+}, \label{e18}
\end{aligned}
\end{equation}
where
\begin{equation}
\left.\frac{\partial f}{\partial M}\right|_{r=r_{+}}=-\frac{2}{\pi r_{+}},\left.\frac{\partial f}{\partial Q}\right|_{r=r_{+}}=\frac{8 Q}{\pi r_{+}^{2}},\left.\frac{\partial f}{\partial r_{+}}\right|_{r=r_{+}}=4 \pi T.\label{e19}
\end{equation}
Bring Eqs.  (\ref{e16}), (\ref{e17}), (\ref{e19}) into Eq.  (\ref{e18}) leads to 
\begin{equation}
d M=\frac{4 Q}{r_{+}} d Q+2 \pi^2 T r_{+} d r_{+}.\label{e20}
\end{equation}
Using the Eqs. (\ref{e4}), (\ref{e15}) and (\ref{e20}), we can obtain the first law of thermodynamics
\begin{equation}
d M=T d S+\varphi d Q.
\end{equation}
In normal phase space, Eq. (\ref{e13}) gives the transmitted energy and charge at a certain time interval
\begin{equation}
d M=\omega(\omega-q \varphi) r_{+}^{2} d t,\quad d Q=q(\omega-q \varphi) r_{+}^{2} d t.\label{e22}
\end{equation}
Bring these results into the Eq. (\ref{e17}), we can obtain the variation of the radius at horizon
\begin{equation}
d r_{+}=\frac{r_{+}}{2 \pi^2 T}(\omega-q \varphi)^{2} d t.
\end{equation}
Hence, the variation of the entropy can be expressed as
\begin{equation}
d S=2 \pi r_{+} d r_{+}=\frac{r_{+}^{2}}{\pi T}(\omega-q \varphi)^{2} d t.
\end{equation}
\begin{figure}[H]
	\centering
	\includegraphics[width=0.7\linewidth]{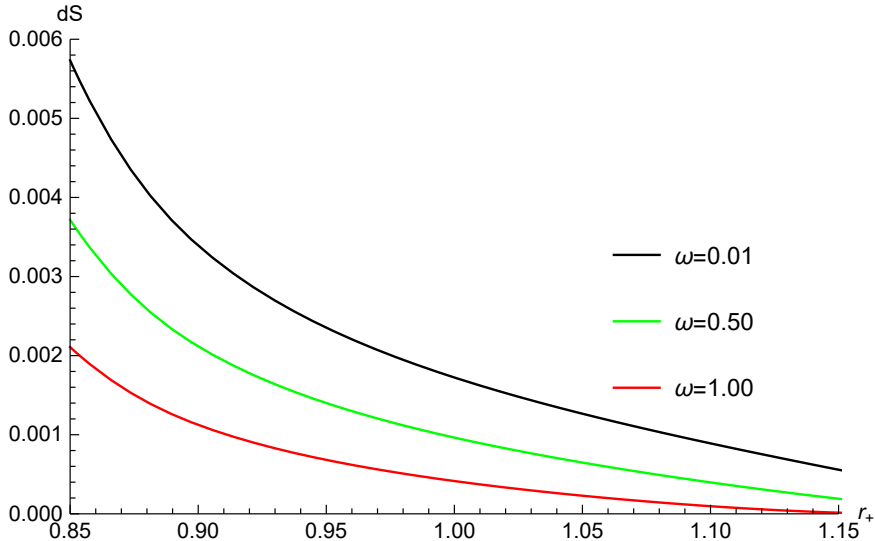}
	\caption{The variation of the $dS$ in normal phase space}
	\label{torus-like-normal-space-dS}
\end{figure}
We use FIG. \ref{torus-like-normal-space-dS} to check whether the second law of thermodynamics in the normal phase space is valid.  We set $M=1.0$, $q=1.0$, the time interval $dt=0.001$, and the cosmological constant $\Lambda=-1.0$. We substitute the above conditions into formula (\ref{e3}). In the calculation of the radius of the black hole's event horizon $r_+$, the maximum value of the black hole's charge is 0.541385 to ensure that the metric represents a black hole instead of a naked singularity. Then, we took $\omega=0.01$ (the black line), $\omega=0.50$ (the green line) and $\omega=1.00$ (the red line) to make this figure. As shown in FIG. \ref{torus-like-normal-space-dS}, we can see that the entropy increases under our assumptions. Therefore, the second law of thermodynamics is not violated.

Then we discuss the validity of the weak cosmic censorship conjecture at the extremal and near-extremal of a black hole in normal phase space. For the torus-like black hole, there is a minimum negative value for the function $f(r)$ with the radial coordinate $r_{0}$. If $f(r_{0})>0$, there is not a horizon. If $f(r_{0}) \leq 0$, there are horizons always. At $r_{0},$ then we can know
\begin{align}
&\left.f\right|_{r=r_{0}} \equiv f_{0}=\delta \leq 0,  \nonumber\\
&\partial_{r}\left.f\right|_{r=r_{0}} \equiv f_{0}^{\prime}=0,\\ &\left(\partial_{r}\right)^{2}\left.f\right|_{r=r_{0}} \equiv f_{0}^{\prime \prime}>0.\nonumber
\end{align}
For extreme black holes, $\delta=0$. Changes in conservative quantities, such as the mass and charge of a black hole, can be written as $(M+dM,Q+dQ)$. After the black hole spreading scalar field, the location of the minimum value and event horizon can be written as $r_ {0} + d r_ {0} $, respectively. Here, we convert the function $f (r) $, marked as $d f (r_ {0}) $. Here are some of the formulas we will use:
\begin{equation}
\left.\frac{\partial f}{\partial M}\right|_{r=r_{0}}=-\frac{2}{\pi r_{0}}, \quad\left.\frac{\partial f}{\partial Q}\right|_{r_{r}-r_{0}}=\frac{8 Q}{\pi r_{0}^{2}},\quad\left.\frac{\partial f}{\partial r_{0}}\right|_{r=r_{0}}=0.\label{e26}
\end{equation}
Using the condition $f_{0}^{\prime}=0$, at $r_{0}+d r_{0}$, $f(r)$ can be expressed as
\begin{align}
\left.f\right|_{r=r_{0}+d r_{0}}&=f(r_{0})+d f(r_{0}) \nonumber\\
&=\delta+[\frac{\partial f(r_{0})}{\partial M} d M+\frac{\partial f(r_{0})}{\partial Q} d Q].\label{e27}
\end{align}
Bring Eqs. (\ref{e13}) and (\ref{e26}) into Eq. (\ref{e27}), we can obtain
\begin{equation}
f(r_{0})+d f(r_{0})=\delta-\frac{2}{\pi r_{0}}\left(\omega-\frac{4 q Q}{r_{+}}\right)\left(\omega-\frac{4 q Q}{r_{0}}\right) r_{+}^{2} d t.\label{e28}
\end{equation}
When it is an extreme black hole, $r_{0}=r_{+}$, $T=0$ and $\delta=0$. So the above equation becomes
\begin{equation}
f(r_{0})+d f(r_{0})=-\frac{2}{\pi}(\omega-q \varphi)^{2} r_{+} d t=0,
\end{equation}
which indicates that the horizon exists in the finial state. The black hole can not be overcharged by the scattering of the scalar filed.

\begin{figure}[H]
	\centering
	\includegraphics[width=0.7\linewidth]{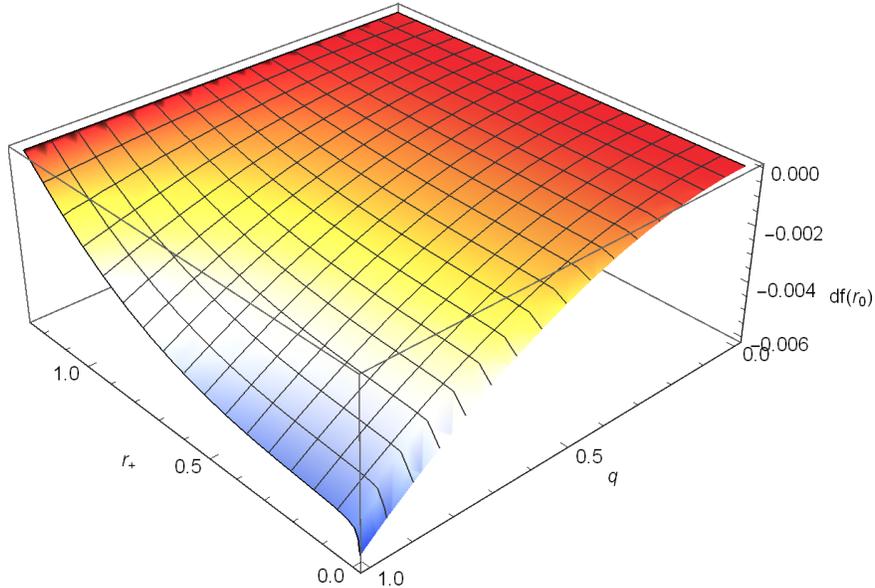}
	\caption{The variation of the $df(r_0)$ in normal phase space}
	\label{torus-like-normal space-df}
\end{figure}

When it is a near-extremal black hole, the Eq. (\ref{e28}) above equation can be regarded as
a quadratic function of $\omega$. When $\omega=2 q Q\left(\frac{1}{r_{+}}+\frac{1}{r_{0}}\right)$, we find an maximum value on the function of $f(r_{0})+d f(r_{0})$
\begin{equation}
[f(r_{0})+d f(r_{0})]_{max}=\delta+\frac{8 q^{2} Q^{2}\left(r_{+}-r_{0}\right)^{2}}{ \pi r_{0}^{3}} dt.
\end{equation}
To gain an intuitive understanding, we plot FIG. \ref{torus-like-normal space-df} in follow
In FIG. \ref{torus-like-normal space-df}, for $M=1.0$, $\omega=0.1$, the time interval $dt=0.001$, $\epsilon=0.001$, and the cosmological constant $\Lambda=-1.0$, there is no region where the value of $df(r_{0})$ is positive. In other word, the function is always negative for the near-extremal black hole in the normal phase space. So the weak cosmic censorship conjecture of the near-extremal black hole is valid in the normal phase space.

\section{The thermodynamics and WCCC of the torus-like black hole in the extended phase space}
Next, we will study the thermodynamics and weak cosmic censorship conjecture in the extended phase space. That is, the cosmological constant $\Lambda$ will be used as the thermodynamic pressure, and the conjugate amount as the thermodynamic volume can be expressed as
\begin{equation}
P=-\frac{\Lambda}{8\pi}, \quad V=\frac{4\pi^{2}r_{+}^{3}}{3}.\label{e31}
\end{equation}
In the extended phase space, the initial state of the black hole is represented by $\left(M, Q, r_{+}, P\right)$, and the final state is $\left(M+d M, Q+d Q, r_{ +}+d r_{+}, P+dP\right)$. The variation in radius can be obtained by measuring the variation in the function $f\left(r_{+}\right)$. For the initial state $\left(M, Q, r_{+}, P\right)$, satisfied
\begin{equation}
f\left(M, Q, r_{+},P\right)=0.\label{e32}
\end{equation}
We assume that the final state of the black hole is still satisfies
\begin{equation}
f\left(M+d M, Q+d Q, r_{+}+d r_{+}, P+d P\right)=0.\label{e33}
\end{equation}
The functions $f\left(M+d M, Q+d Q, r_{+}+d r_{+}, P+d P, a+d a\right)$ and $f\left(M, Q, r_{+}, P, a\right)$ satisfy the following relation
\begin{align}
&  f\left(M+d M, Q+d Q, r_{+}+d r_{+}, P+d P\right)=f\left(M, Q, r_{+}, P\right)\nonumber\\
&  +\left.\frac{\partial f}{\partial M}\right|_{r=r_{+}} d M+\left.\frac{\partial f}{\partial Q}\right|_{r=r_{+}} d Q+\left.\frac{\partial f}{\partial r_{+}}\right|_{r=r_{+}} d r_{+}+\left.\frac{\partial f}{\partial P}\right|_{r=r_{+}} d P,\label{e34}
\end{align}
where
\begin{equation}
\left.\frac{\partial f}{\partial M}\right|_{r=r_{+}}=-\frac{2}{\pi r_{+}},\left.\frac{\partial f}{\partial Q}\right|_{r=r_{+}}=\frac{8 Q}{\pi r_{+}^{2}},\left.\frac{\partial f}{\partial r_{+}}\right|_{r=r_{+}}=4 \pi T,\left.\frac{\partial f}{\partial P}\right|_{r=r_{+}}=\frac{8 \pi r_{+}^{2}}{3}.\label{e35}
\end{equation}
Bring Eqs. (\ref{e32}), (\ref{e33}) and (\ref{e35}) to Eq. (\ref{e34}) leads to
\begin{equation}
d M=2 \pi^{2} r_{+} T d r_{+}+\frac{4 Q}{r_{+}} d Q +\frac{4 \pi^{2} r_{+}^{3}}{3} d P.\label{e36}
\end{equation}
Using the Eqs. (\ref{e31}) and (\ref{e36}), we know the first law of thermodynamics
\begin{equation}
d M=T d S+\varphi dQ +VdP.
\end{equation}
From Eqs. (\ref{e13}), (\ref{e14}), (\ref{e31}) and $\varphi=\frac{4Q}{r_{+}}$, we can obtain the Smarr formula
\begin{equation}
M=2(T S-V P)+\varphi Q.
\end{equation}
Considering that $M$ has a physical meaning of enthalpy in the extended phase space, its relationship with the internal energy of the system is
\begin{equation}
M=U+P V.\label{39}
\end{equation}
Inserting Eq. (\ref{e31}) to Eq. (\ref{39}) yield
\begin{equation}
d M=d(U+P V)=\omega(\omega-q \varphi) r_{+}^{2} d t+\frac{4 \pi^{2} r_{+}^{3}}{3} d P-\frac{\Lambda \pi}{2}r_{+}^{2}d r_{+}.
\end{equation}
Bring these results into the Eq. (\ref{e36}), we can obtain the variation of the radius in horizon
\begin{equation}
d r_{+}=\frac{2(\omega -q \varphi )^{2} r_+}{\pi  (\Lambda  r_+ +4 \pi  T)} dt.
\end{equation}
The variation of the entropy then takes on the form
\begin{equation}
d S=2 \pi^{2} r_{+} d r_{+}=\frac{4\pi(\omega -q \varphi )^{2} r_+^{2}}{(\Lambda  r_+ +4 \pi  T)} dt.
\end{equation}
\begin{figure}[H]
	\centering
	\includegraphics[width=0.7\linewidth]{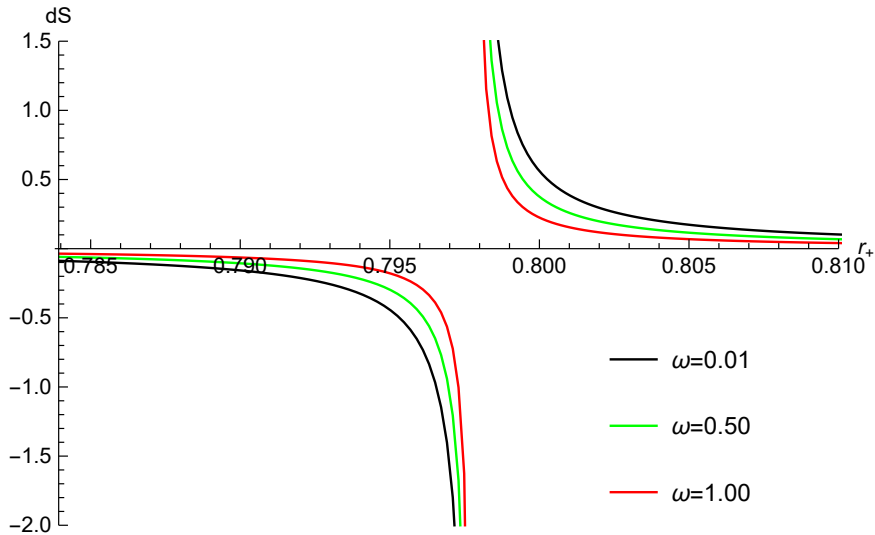}
	\caption{The variation of the $dS$ in extended phase space}
	\label{torus-like-extend space-dS}
\end{figure}
We use FIG. \ref{torus-like-extend space-dS} to illustrate the variation of $dS$ with the horizon radius $r_+$ in the extend phase space. We set $M=1.0$, $q=1.0$, the time interval $dt=0.001$, and the cosmological constant $\Lambda=-1.0$. Then, we took $\omega=0.01$ (the black line), $\omega=0.50$ (the green line), $\omega=1.00$ (the red line) to make this picture. As is shown in FIG. \ref{torus-like-extend space-dS}, We find that the entropy of the black hole increases when the radius of the event horizon $r_+$ is larger than 0.7980 as well as the initial charge $Q$ is no more than 0.5411, which means that at the initial moment the black hole is far away from the conditions needed to become a naked singularity. In this case, the second law of thermodynamics holds. However, when $r_+$ is smaller than 0.7980 as well as the initial charge $Q$ is more than 0.5411, which means that at the initial moment the black hole approaches the conditions needed to become a naked singularity. In this case, the second law of thermodynamics is violated.

Similarly, we can also test the weak cosmic censorship conjecture in the extended phase space. Considering the back reaction, using the Eq. (\ref{e22}), the mass $M$, the charge $Q$, the thermodynamic pressure $P$ and so on will be converted to $(M+d M, Q+d Q, \Lambda+ d \Lambda)$, as the absorption complex scalar field. The change in the function $f(r)$ should satisfy
\begin{align}
&f\left(r_{0}+d r_{0}; M+d M, Q+d Q, P+dP\right)\nonumber\\
&=f(r_{0})+d f(r_{0}) \label{e43}\\
&=\delta+\frac{\partial f(r_{0})}{\partial r_{0}} d r_{0}+\frac{\partial f(r_{0})}{\partial M} d M+\frac{\partial f(r_{0})}{\partial Q} d Q+\frac{\partial f(r_{0})}{\partial P} d P \nonumber\\
&=\delta+\delta_{1}+\delta_{2},\nonumber 
\end{align}
where
\begin{align}
&\delta=f(r_{0},M,Q,P),\nonumber\\
&\delta_{1}=-\frac{2 T}{\pi r_{0}} d S+\frac{8 \pi \left(r_0^3-r_+^3\right)}{3 r_0}dP, \label{e44}\\
&\delta{2}=\frac{8 q Q r_+  (\omega-q \varphi)\left( r_+-r_0\right)}{\pi  r_0^2}dt. \nonumber  
\end{align}
When it is an extreme black hole, $r_{0}=r_{+}$ and $T=0$, we can obtain $\delta=0$, $\delta_{1}=0$ and $\delta_{2}=0$. Hence Eq. (\ref{e43}) can be rewritten as
\begin{equation}
f\left(r_{0}+d r_{0} ; M+d M, Q+d Q, P+dP\right)=0.
\end{equation}
This indicates that the scattering does not change the minimum value. This means that the final state of the extremal black hole is still an extreme black hole, and the weak cosmic censorship conjecture is effective in extending the torus-like black hole of the phase space.

\begin{figure}[H]
	\centering
	\includegraphics[width=0.7\linewidth]{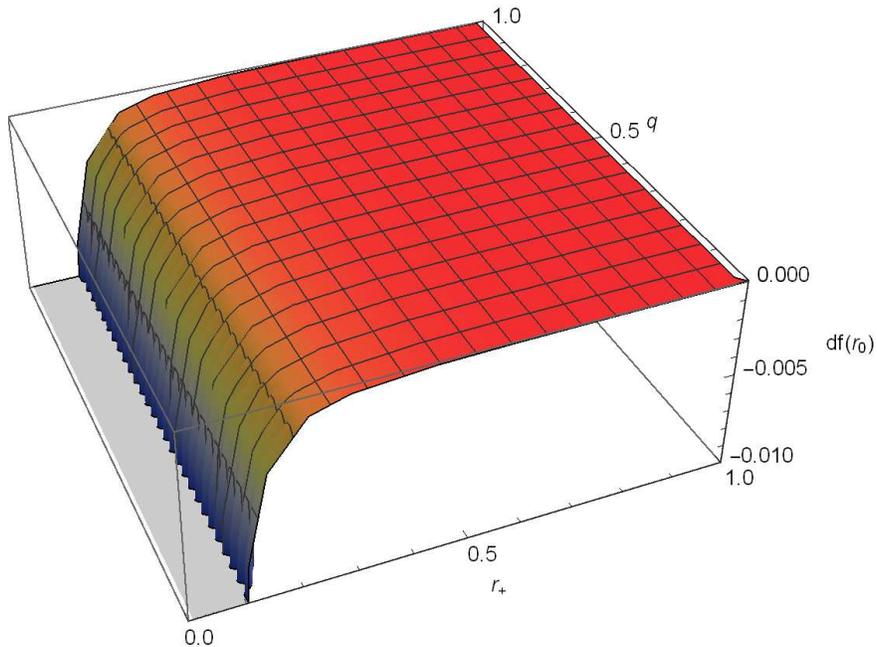}
	\caption{The variation of the $df(r_0)$ in extend phase space}
	\label{torus-like-extend-space-df}
\end{figure}

When it is a near-extremal black hole, the order of the variables becomes important,
$f^{\prime}\left(r_{+}\right)$ is very close to zero. To evaluate the value of the above equation, we can let $r_{+}=r_{0}+\epsilon$, where $0<\epsilon \ll 1$ and the relation $f\left(r_{+}\right)=0$ and $f^{\prime}\left(r_{0}\right)=0$. And then, the Eq. (\ref{e44}) can be written as
\begin{align}
&\delta<0, \nonumber\\
&\delta_{1} =-\frac{f^{\prime \prime}\left(r_{+}\right)}{2 \pi^{2} r_{+}} \epsilon d S+\frac{f^{\prime \prime\prime}\left(r_{+}\right)}{2 \pi^{2} r_{+}^{2}} \epsilon^{2} d S-8 \pi r_+ \epsilon dP+O\left(\epsilon^{3}\right),\\
&\delta_{2} =\frac{8 q Q (\omega -q \varphi )}{\pi  r_+}\epsilon dt +\frac{16  q Q  (\omega -q \varphi )}{\pi  r_+^2}\epsilon ^2dt+O\left(\epsilon^{3}\right).\nonumber
\end{align}
If the initial black hole is near extremal, we have $d S \sim \epsilon$ and $d P \sim \epsilon$. So $\delta 1+\delta 2 \ll \delta$, the final black hole has
\begin{equation}
f\left(r_{0}+d r_{0}, M+d M, Q+d Q, P+dP\right) \approx \delta<0.\label{e48}
\end{equation}
In addition, to know whether $\delta_{1}+\delta_{2}$ is positive, we plot FIG. \ref{torus-like-extend-space-df} in follow
In FIG. \ref{torus-like-extend-space-df}, we illustrate the variation of $df(r_0)$ with the horizon radius $r_+$ and charge $q$ in the extend phase space. We set $M=1.0$, $\omega=0.1$, the time interval $dt=0.001$, $\epsilon=0.001$, and the cosmological constant $\Lambda=-1.0$. From the figure we can see that $df(r_0)$ is less than or equal to zero during the change.

Therefore, Eq. (\ref{e48}) is negative, which indicates that the event horizon exists in the final state. The black hole cannot be overcharged by the scattering of the scalar field.  The weak cosmic censorship conjecture is valid in the near-extremal torus-like black hole.
\section{Conclusion}
In this paper, we obtain a torus-like black hole by scattering a complex scalar field. Using it, we calculate the amount of energy and charge transferred between the scalar field and the black hole over a certain time interval. Then we use it to study the laws of thermodynamics and weak cosmic censorship conjecture in normal phase space and extended phase space.

In the normal phase space, the first and second laws of thermodynamics and weak cosmic censorship conjecture are valid, which are well expressed in FIG. \ref{torus-like-normal-space-dS} and FIG. \ref{torus-like-normal space-df}. For extreme black holes in normal phase space, we find that entropy is infinite, because $T\rightarrow 0$, which means that according to the relationship between entropy and the area of the horizon, the horizon is also infinite. As a result, the singularity is always hidden, which is consistent with the weak cosmic censorship conjecture. For near-extremal black holes in normal phase space, the temperature is not close to zero. The black hole cannot be overcharged by the scattering of the scalar field.

In the extend phase space, the first and weak cosmic censorship conjecture are valid, which is well expressed in FIG. \ref{torus-like-extend-space-df}. However, the second law of thermodynamics is violated in this phase space. We find that the entropy of the black hole increases when the radius of the event horizon $r_+$ is larger than 0.7980 as well as the initial charge $Q$ is no more than 0.5411, which means that at the initial moment the black hole is far away from the conditions needed to become a naked singularity. In this case, the second law of thermodynamics holds. However, when $r_+$ is smaller than 0.7980 as well as the initial charge $Q$ is more than 0.5411, which means that at the initial moment the black hole approaches the conditions needed to become a naked singularity.

Finally, the first law of thermodynamics and the weak cosmic censorship conjecture hold in both normal phase space and extended phase space. However, the validity of the second law of thermodynamics depends on the phase space we choose and the initial conditions of the black hole.

\begin{acknowledgments}
The authors contributed equally to this work. We are grateful to Peng Wang for useful discussions. This work is supported in part by NSFC (Grant No. 11847305). Natural Science Foundation of Chengdu University of TCM (Grants No. ZRYY1729 and ZRQN1656). Discipline Talent Promotion Program of /Xinglin Scholars(Grant No. QNXZ2018050) and the Key Fund Project for Education Department of Sichuan (Grant No. 18ZA0173). Sichuan University Students Platform for Innovation and Entrepreneurship Training Program (Grant No. C2019104639).
\end{acknowledgments}

\end{document}